\documentstyle[12pt]{article}
\newcommand\bea{\begin{eqnarray}}
\newcommand\eea{\end{eqnarray}}

\begin{document}
\thispagestyle{empty}
\bibliographystyle{unsrt}
\setlength{\baselineskip}{18pt}
\parindent 24pt
\vspace{60pt}

\begin{center}{
{\Large {{\bf Quantum decoherence and classical correlations
\\in open systems}} } \vskip 1truecm
A. Isar ${^{(a)}}$\\
{\it Department of Theoretical Physics, Institute of
Physics and Nuclear Engineering,
Bucharest-Magurele, Romania }\\
}
\end{center}

\vskip 1truecm
\begin{abstract}
In the framework of the Lindblad theory for open quantum systems we
determine the degree of quantum decoherence and classical
correlations of a harmonic oscillator interacting with a thermal
bath.  The transition from quantum to classical behaviour of the
considered system is analyzed and it is shown that the classicality
takes place during a finite interval of time. We calculate also the
decoherence time and show that it has the same scale as the time
after which thermal fluctuations become comparable with quantum
fluctuations.
\end{abstract}

(a) e-mail address: isar@theory.nipne.ro

\section{Introduction}

The transition from quantum to classical physics and
classicality of quantum systems continue to be among the
most interesting problems in many fields of physics, for
both conceptual and experimental reasons
\cite{gi96,pa01,zu03}. Two conditions are essential for
the classicality of a quantum system \cite{mo90}: a)
quantum decoherence (QD), that means the irreversible,
uncontrollable and persistent formation of a quantum
correlation (entanglement) of the system with its
environment \cite{ali}, expressed by the damping of the
coherences present in the quantum state of the system,
when the off-diagonal elements of the density matrix of
the system decay below a certain level, so that this
density matrix becomes approximately diagonal and b)
classical correlations (CC), expressed by the fact that
the Wigner function of the quantum system has a peak
which follows the classical equations of motion in phase
space with a good degree of approximation, that is the
quantum state becomes peaked along a classical
trajectory. The necessity and sufficiency of both QD and
CC as conditions of classicality are still a subject of
debate. Both these conditions do not have an universal
character, so that they are not necessary for all
physical models. An important role in this discussion
plays the temperature of the environment and therefore it
is worth to take into account the differences between low
and high temperature regimes. For example, purely
classical systems at very high temperatures are described
by a classical Fokker-Planck equation which does not
follow any trajectory in phase space (for very small
kinetic energy, compared to the thermal energy, when the
probability distribution becomes essentially independent
of momentum), so that in this case CC are not necessary.
Likewise, one can have a classical behaviour if the
coherences are negligible, without having strong CC (for
example, in the case of a classical gas at finite
temperature) and the lack of strong correlations between
the coordinate and its canonical momentum does not
necessarily mean that the system is quantum. On the other
hand, the condition of CC is not sufficient for a system
to become classical -- although the Wigner function can
show a sharp correlation in phase space, the quantum
coherence never vanishes for a closed system which has a
unitary evolution. Likewise, in the low temperature
quantum regime one can observe strong CC. For example, in
the case of a purely damped quantum harmonic oscillator
(at zero temperature), the initial coherent states remain
coherent and perfectly follow classical trajectories of a
damped oscillator, but CC are not sufficient for
classicality.

In the last two decades it has became more and more clear
that the classicality is an emergent property of open
quantum systems, since both main features of this process
-- QD and CC -- strongly depend on the interaction
between the system and its external environment
\cite{zu03,pa93,zu91}. The main purpose of this work is
to study QD and CC for a harmonic oscillator interacting
with an environment in the framework of the Lindblad
theory for open quantum systems. We determine the degree
of QD and CC and the possibility of simultaneous
realization of QD and CC for a system consisting of a
harmonic oscillator in a thermal bath. It is found that
the system manifests a QD which increases with time and
temperature, whereas CC are less and less strong with
increasing time and temperature.

\section{Lindblad master equation for the harmonic
oscillator in coordinate and Wigner representation}

The irreversible time evolution of an open system is described by
the following general quantum Markovian master equation for the
density operator $\rho(t)$ \cite{l1}: \bea{d \rho(t)\over
dt}=-{i\over\hbar}[ H, \rho(t)]+{1\over 2\hbar} \sum_{j}([  V_{j}
\rho(t), V_{j}^\dagger ]+[ V_{j}, \rho(t) V_{j}^\dagger
]).\label{lineq}\eea $H$ is the Hamiltonian of the system and
$V_{j},$ $ V_{j}^\dagger $ are operators on the Hilbert space of
$H$, which model the environment. In order to obtain, for the damped
quantum harmonic oscillator, equations of motion as close as
possible to the classical ones, the two possible operators $V_{1}$
and $ V_{2}$ are taken as linear polynomials in coordinate $q$ and
momentum $p$ \cite{ss,rev} and the harmonic oscillator Hamiltonian
$H$ is chosen of the general quadratic form \bea H=H_{0}+{\mu\over
2}(qp+pq), ~~~  H_{0}={1\over 2m}p^2+{m\omega^2\over 2}  q^2.
\label{ham} \eea With these choices the master equation
(\ref{lineq}) takes the following form \cite{ss,rev}: \bea {d \rho
\over dt}=-{i\over \hbar}[ H_{0}, \rho]- {i\over 2\hbar}(\lambda
+\mu) [  q, \rho p+ p \rho]+{i\over 2\hbar}(\lambda -\mu)[  p,
\rho   q+  q \rho]  \nonumber\\
-{D_{pp}\over {\hbar}^2}[ q,[  q, \rho]]-{D_{qq}\over {\hbar}^2} [
p,[  p, \rho]]+{D_{pq}\over {\hbar}^2}([ q,[  p, \rho]]+ [ p,[ q,
\rho]]). ~~~~\label{mast}   \eea In the particular case when the
asymptotic state is a Gibbs state $\rho_G(\infty)=e^{-{  H_0\over
kT}}/{\rm Tr}e^{-{  H_0\over kT}},$ the quantum diffusion
coefficients $D_{pp},D_{qq},$ $D_{pq}$ and the dissipation constant
$\lambda$ satisfy the relations \cite{ss,rev} \bea
D_{pp}={\lambda+\mu\over 2}\hbar m\omega\coth{\hbar\omega\over 2kT},
~~D_{qq}={\lambda-\mu\over 2}{\hbar\over
m\omega}\coth{\hbar\omega\over 2kT}, ~~D_{pq}=0, \label{coegib} \eea
where $T$ is the temperature of the thermal bath.

In the Markovian regime the harmonic oscillator master
equation which satisfies the complete positivity
condition cannot satisfy simultaneously the translational
invariance and the detailed balance (which assures an
asymptotic approach to the canonical thermal equilibrium
state). The necessary and sufficient condition for
translational invariance is $\lambda=\mu$ \cite{ss,rev}.
In this case the equations of motion for the expectation
values of coordinate and momentum are exactly the same as
the classical ones. If $\lambda\neq \mu,$ then we violate
translational invariance, but we keep the canonical
equilibrium state.

The asymptotic values $\sigma_{qq}(\infty),
\sigma_{pp}(\infty),\sigma_{pq}(\infty)$ do not depend on
the initial values
$\sigma_{qq}(0),\sigma_{pp}(0),\sigma_{pq}(0)$ and in the
case of a thermal bath with coefficients (\ref{coegib}),
they reduce to \cite{ss,rev} \bea
\sigma_{qq}(\infty)={\hbar\over
2m\omega}\coth{\hbar\omega\over 2kT},
~~\sigma_{pp}(\infty)={\hbar m\omega\over
2}\coth{\hbar\omega\over 2kT}, ~~\sigma_{pq}(\infty)=0.
\label{varinf}\eea

In the following, we consider a general temperature $T,$ but we
should stress that the Lindblad theory is obtained in the Markov
approximation, which holds for high temperatures of the environment.
At the same time, the semigroup dynamics of the density operator
which must hold for a quantum Markovian process is valid only for
the weak-coupling regime, with the damping $\lambda$ obeying the
inequality $\lambda\ll\omega.$

We consider a harmonic oscillator with an initial Gaussian wave
function \bea \Psi(q)=({1\over 2\pi\sigma_{qq}(0)})^{1\over
4}\exp[-{1\over 4\sigma_{qq}(0)}
(1-{2i\over\hbar}\sigma_{pq}(0))(q-\sigma_q(0))^2+{i\over
\hbar}\sigma_p(0)q], \label{ccs}\eea where $\sigma_{qq}(0)$ is the
initial spread, $\sigma_{pq}(0)$ the initial covariance, and
$\sigma_q(0)$ and $\sigma_p(0)$ are the initial averaged position
and momentum of the wave packet. The initial state (\ref{ccs})
represents a correlated coherent state (squeezed coherent state)
\cite{dodkur} with the variances and covariance of coordinate and
momentum \bea \sigma_{qq}(0)={\hbar\delta\over 2m\omega},~~
\sigma_{pp}(0)={\hbar m\omega\over 2\delta(1-r^2)},~~
\sigma_{pq}(0)={\hbar r\over 2\sqrt{1-r^2}}. \label{inw}\eea Here,
$\delta$ is the squeezing parameter which measures the spread in the
initial Gaussian packet and $r,$ with $|r|<1$ is the correlation
coefficient at time $t=0.$ The initial values (\ref{inw}) correspond
to a minimum uncertainty state, since they fulfil the generalized
uncertainty relation \bea
\sigma_{qq}(0)\sigma_{pp}(0)-\sigma_{pq}^2(0) ={\hbar^2\over
4}.\label{gen0}\eea For $\delta=1$ and $r=0$ the correlated coherent
state becomes a Glauber coherent state.

From Eq. (\ref{mast}) we derive the evolution equation in coordinate
representation: \bea {\partial\rho\over\partial t}={i\hbar\over
2m}({\partial^2\over\partial q^2}- {\partial^2\over\partial
q'^2})\rho-{im\omega^2\over
2\hbar}(q^2-q'^2)\rho\nonumber\\
-{1\over 2}(\lambda+\mu)(q-q')({\partial\over\partial
q}-{\partial\over\partial q'})\rho+{1\over
2}(\lambda-\mu)[(q+q')({\partial\over\partial
q}+{\partial\over\partial
q'})+2]\rho  \nonumber\\
-{D_{pp}\over\hbar^2}(q-q')^2\rho+D_{qq}({\partial\over\partial
q}+{\partial\over \partial q'})^2\rho
-{2iD_{pq}\hbar}(q-q')( {\partial\over\partial
q}+{\partial\over\partial q'})\rho.\label{cooreq}\eea For
the case of a thermal bath with coefficients
(\ref{coegib}) the Wigner distribution function
$W(q,p,t)$ satisfies the following Fokker-Planck-type
equation: \bea {\partial W\over\partial t}= -{p\over
m}{\partial W\over\partial q} +m\omega^2 q{\partial
W\over\partial p} +(\lambda+\mu){\partial\over\partial
p}(pW)
+(\lambda-\mu){\partial\over\partial q}(qW) \nonumber \\
+{\hbar\over 2}\coth{\hbar\omega\over
2kT}[(\lambda+\mu)m\omega{\partial^2 W\over\partial p^2}
+{\lambda-\mu\over m\omega}{\partial^2 W\over\partial
q^2}].~~~~~~~~~~~~~~~~~~~ \label{wigeq}\eea The first two
terms on the right-hand side of both these equations
generate a purely unitary evolution. They give the usual
Liouvillian evolution. The third and forth terms are the
dissipative terms and have a damping effect (exchange of
energy with environment). The last two are noise
(diffusive) terms and produce fluctuation effects in the
evolution of the system. They promote diffusion in
momentum $p$ and coordinate $q$ and generate decoherence
in coordinate and momentum, respectively. In the high
temperature limit, quantum Fokker-Planck equation
(\ref{wigeq}) with coefficients (\ref{coegib}) becomes
classical Kramers equation ($D_{pp}\to 2m\lambda kT$ for
$\lambda=\mu$).

The density matrix solution of Eq. (\ref{cooreq}) has the general
form of Gaussian density matrices \bea <q|\rho(t)|q'>=({1\over
2\pi\sigma_{qq}(t)})^{1\over 2} \exp[-{1\over
2\sigma_{qq}(t)}({q+q'\over
2}-\sigma_q(t))^2\nonumber\\
-{\sigma(t)\over 2\hbar^2\sigma_{qq}(t)}(q-q')^2
+{i\sigma_{pq}(t)\over \hbar\sigma_{qq}(t)}({q+q'\over
2}-\sigma_q(t))(q-q')+{i\over
\hbar}\sigma_p(t)(q-q')],\label{densol} \eea where
$\sigma(t)\equiv\sigma_{qq}(t)\sigma_{pp}(t)-\sigma_{pq}^2(t)$
is the Schr\"odinger generalized uncertainty function
\cite{unc} ($\sigma_{qq}$ and $\sigma_{pp}$ denote the
dispersion (variance) of the coordinate and momentum,
respectively, and $\sigma_{pq}$ denotes the correlation
(covariance) of the coordinate and momentum).

For an initial Gaussian Wigner function (corresponding to a
correlated coherent state (\ref{ccs})) the solution of Eq.
(\ref{wigeq}) is \bea W(q,p,t)={1\over 2\pi\sqrt{\sigma(t)}}
\exp\{-{1\over 2\sigma(t)}[\sigma_{pp}(t)(q-\sigma_q(t))^2+
\sigma_{qq}(t)(p-\sigma_p(t))^2\nonumber\\
-2\sigma_{pq}(t)(q-\sigma_q(t))(p-\sigma_p(t))]
\}.\label{wig} \eea

In the case of a thermal bath we obtain the following steady state
solution for $t\to\infty$ (we denote
$\epsilon\equiv{\hbar\omega\over 2kT}$): \bea
<q|\rho(\infty)|q'>=({m\omega\over \pi\hbar\coth\epsilon})^{1\over
2}\exp\{-{m\omega\over 4\hbar}[{(q+q')^2\over\coth\epsilon}+
(q-q')^2\coth\epsilon]\}.\label{dinf}\eea In the long time limit we
have also \bea W_{\infty}(q,p)={1\over \pi\hbar\coth
\epsilon}\exp\{-{1\over \hbar\coth\epsilon}[m\omega q^2+{p^2\over
m\omega}] \}.\label{wiginf} \eea Stationary solutions to the
evolution equations obtained in the long time limit are possible as
a result of a balance between the wave packet spreading induced by
the Hamiltonian and the localizing effect of the Lindblad operators.

\section{Quantum decoherence and classical correlations}

As we already stated, one considers that two conditions
have to be satisfied in order that a system could be
considered as classical. The {\it first} condition
requires that the system should be in one of relatively
permanent states -- states that are least affected by the
interaction of the system with the environment -- and the
interference between different states should be
negligible. This implies the destruction of off-diagonal
elements representing coherences between quantum states
in the density matrix, which is the QD phenomenon. The
loss of coherence can be achieved by introducing an
interaction between the system and environment: an
initial pure state with a density matrix which contains
nonzero off-diagonal terms can non-unitarily evolve into
a final mixed state with a diagonal density matrix during
the interaction with the environment, like in classical
statistical mechanics.

The {\it second} condition requires that the system
should have, with a good approximation, an evolution
according to classical laws. This implies that the Wigner
distribution function has a peak along a classical
trajectory, that means there exist CC between the
canonical variables of coordinate and momentum. Of
course, the correlation between the canonical variables,
necessary to obtain a classical limit, should not violate
Heisenberg uncertainty principle, i.e. the position and
momentum should take reasonably sharp values, to a degree
in concordance with the uncertainty principle. This is
possible, because the density matrix does not diagonalize
exactly in position, but with a non-zero width.

Using new variables $\Sigma=(q+q')/2$ and $\Delta=q-q',$ the density
matrix (\ref{densol}) can be rewritten as \bea
\rho(\Sigma,\Delta,t)=\sqrt{\alpha\over \pi}\exp[-\alpha\Sigma^2
-\gamma\Delta^2+i\beta\Sigma\Delta+2\alpha\sigma_q(t)\Sigma
+i({\sigma_p(t)\over\hbar}-
\beta\sigma_q(t))\Delta-\alpha\sigma_q^2(t)],\label{ccd3}\eea with
the abbreviations \bea \alpha={1\over
2\sigma_{qq}(t)},~~\gamma={\sigma(t)\over 2\hbar^2
\sigma_{qq}(t)},~~
\beta={\sigma_{pq}(t)\over\hbar\sigma_{qq}(t)}\label{ccd4}\eea and
the Wigner transform of the density matrix (\ref{ccd3}) is \bea
W(q,p,t)={1\over
2\pi\hbar}\sqrt{\alpha\over\gamma}\exp\{-{[\hbar\beta
(q-\sigma_q(t))-(p-\sigma_p(t))]^2\over 4\hbar^2\gamma}-\alpha
(q-\sigma_q(t))^2\}.\label{wigc} \eea

a) {\it Degree of quantum decoherence (QD)}

The representation-independent measure of the degree of QD
\cite{mo90} is given by the ratio of the dispersion
$1/\sqrt{2\gamma}$ of the off-diagonal element $\rho(0,\Delta,t)$ to
the dispersion $\sqrt{2/\alpha}$ of the diagonal element
$\rho(\Sigma,0,t):$ \bea \delta_{QD}={1\over 2}\sqrt{\alpha\over
\gamma},\label{qdec}\eea which in our case gives \bea
\delta_{QD}(t)={\hbar\over 2\sqrt{\sigma(t)}}.\eea

The finite temperature Schr\"odinger generalized uncertainty
function, calculated in Ref. \cite{unc}, has the expression
\bea\sigma(t)={\hbar^2\over 4}\{e^{-4\lambda
t}[1-(\delta+{1\over\delta(1-r^2)})\coth\epsilon+\coth^2\epsilon]
\nonumber\\
+e^{-2\lambda
t}\coth\epsilon[(\delta+{1\over\delta(1-r^2)}
-2\coth\epsilon){\omega^2-\mu^2\cos(2\Omega
t)\over\Omega^2}\nonumber \\
+(\delta-{1\over\delta(1-r^2)}){\mu \sin(2\Omega
t)\over\Omega}+{2r\mu\omega(1-\cos(2\Omega
t))\over\Omega^2\sqrt{1-r^2}}]+\coth^2\epsilon\}.\label{sunc}\eea In
the limit of long times Eq. (\ref{sunc}) yields \bea
\sigma(\infty)={\hbar^2\over 4}\coth^2\epsilon,\eea so that we
obtain \bea \delta_{QD}(\infty)=\tanh{\hbar\omega\over 2kT},\eea
which for high $T$ becomes \bea\delta_{QD}(\infty)={\hbar\omega\over
2kT}.\eea

We see that $\delta_{QD}$ decreases, and therefore QD increases,
with temperature, i.e. the density matrix becomes more and more
diagonal at higher $T$ and the contributions of the off-diagonal
elements get smaller and smaller. At the same time the degree of
purity decreases and the degree of mixedness increases with $T.$
$\delta_{QD}<1$ for $T\neq 0,$ while for $T=0$ the asymptotic
(final) state is pure and $\delta_{QD}$ reaches its initial maximum
value 1. $\delta_{QD}= 0$ when the quantum coherence is completely
lost. So, when $\delta_{QD}= 1$ there is no QD and only if
$\delta_{QD}<1,$ there is a significant degree of QD, when the
magnitude of the elements of the density matrix in the position
basis are peaked preferentially along the diagonal $q=q'.$ When
$\delta_{QD}\ll 1,$ we have a strong QD.

b) {\it Degree of classical correlations (CC)}

In defining the degree of CC, the form of the Wigner function is
essential, but not its position around $\sigma_q(t)$ and
$\sigma_p(t).$ Consequently, for simplicity we consider zero values
for the initial expectations values of the coordinate and momentum
and the expression (\ref{wigc}) of the Wigner function becomes \bea
W(q,p,t)={1\over
2\pi\hbar}\sqrt{\alpha\over\gamma}\exp[-{(\hbar\beta q-p)^2\over
4\hbar^2\gamma}-\alpha q^2].\label{wigcs} \eea As a measure of the
degree of CC we take the relative sharpness of this peak in the
phase space determined from the dispersion $\hbar\sqrt{2\gamma}$ in
$p$ in Eq. (\ref{wigcs}) and the magnitude of the average of $p$
($p_0=\hbar\beta q $) \cite{mo90}: \bea
\delta_{CC}={2\sqrt{\alpha\gamma}\over|\beta|},\label{cor}\eea where
we identified $q$ as the dispersion $1/\sqrt{2\alpha}$ of $q.$
$\delta_{CC}$ is a good measure of the "squeezing" of the Wigner
function in phase space \cite{mo90}: in the state (\ref{wigcs}),
more "squeezed" is the Wigner function, more strongly established
are CC.

For our case, we obtain \bea \delta_{CC}(t)={\sqrt{\sigma(t)}\over
|\sigma_{pq}(t)|},\eea where $\sigma(t)$ is given by Eq.
(\ref{sunc}) and $\sigma_{pq}(t)$ can be calculated using formulas
given in Refs. \cite{ss,rev}: \bea\sigma_{pq}(t)={\hbar\over
4\Omega^2}e^{-2\lambda
t}\{[\mu\omega(2\coth\epsilon-\delta-{1\over\delta(1-r^2)})
-{2\omega^2r\over\sqrt{1-r^2}}]\cos(2\Omega t)\nonumber \\
+\omega\Omega(\delta-{1\over\delta(1-r^2)})\sin(2\Omega
t)+\mu\omega(\delta+{1\over\delta(1-r^2)}-2\coth\epsilon)+{2\mu^2r
\over\sqrt{1-r^2}}\}.\label{pqvar}\eea When $\delta_{CC}$
is of order of unity, we have a significant degree of
classical correlations. The condition of strong CC is
$\delta_{CC}\ll 1,$ which assures a very sharp peak in
phase space. Since $\sigma_{pq}(\infty)=0,$ in the case
of an asymptotic Gibbs state, we get
$\delta_{CC}(\infty)\to\infty,$ so that our expression
shows no CC at $t\to\infty.$

c) {\it Discussion with Gaussian density matrix and Wigner function}

If the initial wave function is Gaussian, then the density matrix
(\ref{densol}) and the Wigner function (\ref{wig}) remain Gaussian
for all times (with time-dependent parameters which determine their
amplitude and spread) and centered along the trajectory given by the
solutions of the dissipative equations of motion. This trajectory is
exactly classical for $\lambda=\mu$ and only approximately classical
for not large $\lambda-\mu.$

The degree of QD has an evolution which shows that in general QD
increases with time and temperature. The degree of CC has a more
complicated evolution, but the general tendency is that CC are less
and less strong with increasing time and temperature.
$\delta_{QD}<1$ and $\delta_{CC}$ is of the order of unity for a
long enough interval of time, so that we can say that the considered
system interacting with the thermal bath manifests both QD and CC
and a true quantum to classical transition takes place. Dissipation
promotes quantum coherences, whereas fluctuation (diffusion) reduces
coherences and promotes QD. The balance of dissipation and
fluctuation determines the final equilibrium value of $\delta_{QD}.$
The quantum system starts as a pure state, with a Wigner function
well localized in phase space (Gaussian form). This state evolves
approximately following the classical trajectory (Liouville flow) in
phase space and becomes a quantum mixed state during the
irreversible process of QD.

d) {\it Decoherence time}

In order to obtain the decoherence
time, we consider the coefficient $\gamma$ (\ref{ccd4}), which
measures the contribution of non-diagonal terms in the density
matrix (\ref{ccd3}). For short times ($\lambda t\ll 1, \Omega t\ll
1$), we have: \bea \gamma(t)=-{m\omega\over
4\hbar\delta}\{1+2[\lambda(\delta+{r^2\over\delta(1-r^2)})\coth\epsilon
+\mu(\delta-{r^2\over\delta(1-r^2)})\coth\epsilon-\lambda-\mu-{\omega
r\over\delta\sqrt{1-r^2}}]t\}.\label{td}\eea From here we obtain
that quantum coherences in the density matrix decay exponentially at
a rate given by \bea
2[\lambda(\delta+{r^2\over\delta(1-r^2)})\coth\epsilon
+\mu(\delta-{r^2\over\delta(1-r^2)})\coth\epsilon-\lambda-\mu-{\omega
r\over\delta\sqrt{1-r^2}}]\eea and then the decoherence time scale
is \bea t_{deco}={1\over
2[\lambda(\delta+{r^2\over\delta(1-r^2)})\coth\epsilon
+\mu(\delta-{r^2\over\delta(1-r^2)})\coth\epsilon-\lambda-\mu-{\omega
r\over\delta\sqrt{1-r^2}}]}.\label{tdeco1}\eea The decoherence time
depends on the temperature $T$ and the coupling $\lambda$
(dissipation coefficient) between the system and environment
(through the diffusion coefficient $D_{pp}$), on the squeezing
parameter $\delta$ that measures the spread in the initial Gaussian
packet and on the initial correlation coefficient $r.$ We notice
that the decoherence time is decreasing with increasing dissipation,
temperature and squeezing.

For $r=0$ we obtain:\bea t_{deco}={1\over
2(\lambda+\mu)(\delta\coth\epsilon-1)}\label{tdeco2}\eea and at
temperature $T=0$ (then we have to take $\mu=0$), this becomes \bea
t_{deco}={1\over 2\lambda(\delta-1)}.\eea We see that when the
initial state is the usual coherent state $(\delta=1),$ then the
decoherence time tends to infinity. This corresponds to the fact
that for $T=0$ and $\delta=1$ the coefficient $\gamma$ is constant
in time, so that the decoherence process does not occur in this
case.

At high temperature, expression
(\ref{tdeco1}) becomes (we denote $\tau\equiv
{1\over\epsilon}$)
\bea t_{deco}={1\over
2[\lambda(\delta+{r^2\over\delta(1-r^2)})
+\mu(\delta-{r^2\over\delta(1-r^2)})]\tau}.\eea If, in addition
$r=0,$ then we obtain \bea t_{deco}={\hbar\omega\over
4(\lambda+\mu)\delta kT}.\eea

In Ref. \cite{unc} we determined the time $t_d$ when thermal
fluctuations become comparable with quantum fluctuations: \bea
t_d={1\over 2[\lambda
(\delta+{1\over\delta(1-r^2)})\coth\epsilon+\mu(\delta-{1\over\delta(1-r^2)})
\coth\epsilon-2\lambda]}.\label{t2}\eea At high temperature,
expression (\ref{t2})
becomes \bea t_d={1\over 2\tau[\lambda
(\delta+{1\over\delta(1-r^2)})+\mu(\delta-{1\over\delta(1-r^2)})]}.\eea
As expected, the decoherence time $t_{deco}$ has the same scale as the
time $t_d$ after which thermal fluctuations become
comparable with quantum fluctuations.

We can assert that in the considered case classicality is a
temporary phenomenon, which takes place only at some stages of the
dynamical evolution, during a definite interval of time \cite{deco}.
Due to the dissipative nature of evolution, the approximately
deterministic evolution is no more valid for very large times, when
the localization of the system is affected by the spreading of the
wave packet and of the Wigner distribution function.

\end{document}